# Self-Charged Graphene Battery Harvests Electricity from Thermal Energy of the Environment

Zihan Xu[1†*], Guoan Tai[1,3†], Yungang Zhou[2], Fei Gao[2], and Kin Hung Wong[1]

[1] Department of Applied Physics and Materials Research Centre, The Hong Kong Polytechnic University, Hung Hom, Kowloon, Hong Kong SAR, China

[2] Pacific Northwest National Laboratory, P.O. Box 999, Richland, Washington 99352, USA

[3] The State Key Laboratory of Mechanics and Control of Mechanical Structures, Nanjing University of Aeronautics and Astronautics, 29 Yudao St., Nanjing, China

[†]These authors contributed equally to this work.

*To whom correspondence should be addressed. E-mail: zihan.xuu@gmail.com

**The energy of ionic thermal motion presents universally, which is as high as 4 kJ·kg$^{-1}$·K$^{-1}$ in aqueous solution, where thermal velocity of ions is in the order of hundreds of meters per second at room temperature[1,2]. Moreover, the thermal velocity of ions can be maintained by the external environment, which means it is limitless. However, little study has been reported on converting the ionic thermal energy into electricity. Here we present a graphene device with asymmetric electrodes configuration**

**to capture such ionic thermal energy and convert it into electricity. An output voltage around 0.35 V was generated when the device was dipped into saturated $CuCl_2$ solution, in which this value lasted over twenty days. A positive correlation between the open-circuit voltage and the temperature, as well as the cation concentration, was observed. Furthermore, we demonstrated that this finding is of practical value by lighting a commercial light-emitting diode up with six of such graphene devices connected in series. A typical device showed that the power density can be as high as 70 KW/Kg. This finding provides a new way to understand the behavior of graphene at molecular scale and represents a huge breakthrough for the research of self-powered technology. Moreover, the finding will benefit quite a few applications, such as artificial organs, clean renewable energy and portable electronics, which are longing for novel battery technology.**

Experiment has been reported on collecting electricity from water flow by graphene[3]. Here we designed a device (Fig. 1a) based on graphene, a two-dimensional material exhibiting extremely high intrinsic carrier mobility and large surface-to-volume ratio[4-8], to collect electricity from the thermal motion of ions instead of flows of the solutions.

The open-circuit voltage ($V_{oc}$) of a device with Au-Ag electrodes in 5.56M $CuCl_2$ solution was typically up to 0.45 V (Fig 1b); the output voltage ($V_{op}$) generated by this device loaded a 220 kΩ resistor was around 0.35 V, which could be maintained about 25 days (Fig. 1b). The measurement was taken continuously and the device was kept still in the open area in the lab.

In the second week, the value was undulate due to the air conditioner out of work during the Christmas holiday. The value dropped to about 40 mV after about one month. The damge of the electrodes deposited on graphene was not observed when the device was taken out of the solution. But it was found that graphene grains had fallen off from $SiO_2$ (300 nm)/Si substrate after immersed in the solution for 33 days, which was confirmed by Raman spectrum. (Supplementary, Fig. 4). This continuous measurement was repeated on another sample with graphite-Siliver as electrodes over 7 days (Supplementary ,Figure 5). Six other samples' $V_{oc}$ were also measured, which vary from 0.36 V to 0.51 V (Supplementary Fig. 6).

The dependence of $V_{oc}$ on temperature was investigated (Fig. 2a). 4 M $CuCl_2$ solution was used instead of saturated one to avoid the precipitation of $CuCl_2$ at low temperatures. As can be seen from Fig. 2a, $V_{oc}$ rose with the increase of temperature, and a positive correlation between $V_{oc}$ and temperature was observed. To further verify the relationship between the performance of our device and the velocity of $Cu^{2+}$, we tried ultrasound to treat the $CuCl_2$ solution to increase the velocity of $Cu^{2+}$. The result (Fig. 2b) showed that the $V_{op}$ was increased a little when the ultrasound was on and the value decreased when it was off. The results echoed the $V_{oc}$-temperature experiment.

Furthermore, the effect of $CuCl_2$ solution concentration on the device performance was also investigated (Fig. 2c). It can be seen that the $V_{oc}$ produced by saturated $CuCl_2$ (5.56 M) was about 10 % higher than the dilute one (1 M). From our experiments, a positive correlation between $V_{oc}$ and $[Cu^{2+}]$ was also observed.

Based on electrical double layer, the ions closest to the surface of graphene are cations,

which means that the performance of the graphene battery will be influenced by the cations directly. Thus, we investigated the effect of different cations, such as $Na^+$, $Li^+$, $Co^{2+}$ and $Ni^{2+}$ (Fig. 2d). The results showed that $Cu^{2+}$ can induce the highest $V_{oc}$ compared with other cations. Physiological saline was also used to produce electricity (Supplementary Fig. 7). The result showed that even low concentration of ions can still be used to produce electricity by this kind of device. Other anion solutions such as $CuBr_2$, $Cu(NO_3)_2$ and $CuSO_4$ were also investigated (Fig. 2e). The results showed that the $V_{oc}$ has a negative correlation with the anion radius. That was because when the radius of anion increased, the anion layer will decrease the number of ions which can reach the surface and decrease the total energy that was transported to the surface of graphene.

Our experiments showed that monolayer graphene have the excellent ability to harvest the energy of thermal ions and convert them into electricity (Fig. 2f). We believe that was because the absorbed energy will be exhausted between the layers of graphene by interaction between them when the number of layer is big than two. Thus, we reasonablly predicted that all single-atom-layer materials should have this kind of effect. To obtain such an effect, some need low temperature and others require high temperature. We excluded the possibility of chemical reaction using three groups of control experiments: (1) device with Au-Ag as electrodes in 2 M NaCl solution; (2) device with graphite as electrodes in saturated NaCl (6.1 M) and deionized water. These two experiments were performed to exclude the possibility of chemical reaction between graphene and salt solution, and between solution and electrodes; (3) The third group was performed with high purity gold wires (4N) instead of silver coated copper wire, and excluded the possibility of chemical reaction between conductive wire and solution. (see discussion on the possibility of chemical reaction in Supplementary Information).

Serious structural distortion of graphene was identified by Raman spectroscopy after it was wet by $CuCl_2$ solution (Fig. 2g). By comparison with Raman spectrum of pure graphene (bottom, Fig. 2g), three observations were made: (1) A new $G^-$ band appeared at ~1530 $cm^{-1}$ (middle, Fig. 2g) due to partial bond distortion caused by the bombardment of $Cu^{2+}$ ions[9-12]. Weak $G^-$ bands were observed in graphene wet by other electrolytes (Supplementary Fig. 9). (2) Raman upshifts of G (16 $cm^{-1}$) and G' (8 $cm^{-1}$) bands were observed. (3) A decrease of the $I_{G'}/I_G$ ratio was observed, which can be attributed to the increased absolute value of the Fermi level of graphene[8]. Both (2) and (3) originated from the change of carrier concentration in graphene[8]. The upshifts of the G and G' bands show that electrons were missing in graphene when it was wet by $CuCl_2$ solution. After washing the sample with distilled water, Raman spectrum was restored to the original state (top trace, Fig. 2g).

A thermal ion-graphene interaction mechanism is proposed here to interpret the experimental results. For simplicity, we only consider the interaction between an effective cation ({cation}) and graphene. Here the {cation} can be regarded as a set of $n$ cations. There are two processes involved for the electricity generation: (1) an electron is emitted by the interaction between graphene and the {cation}; (2) the emitted electron flows across the graphene plane to the electrode instead of being transferred to the {cation}. A typical physical model is shown in Fig. 3a. In this model, we define

$$E_{\{cation\},i} - E_{\{cation\},r} > E_{min,1e} = \Phi_{gr} \tag{1}$$

where $E_{\{cation\},i}$, $E_{\{cation\},r}$, $E_{min,1e}$ and $\Phi_{gr}$ are the kinetic energy of the incoming {cation}, the kinetic energy of the rebounded {cation}, the minimum energy required to release a delocalized electron from graphene, and work function of graphene, respectively. $E_{\{cation\},i} = (1/2)mv_i^2$, where $m$ is the mass of the {cation} and $v_i$ is its initial speed, which is determined by the Maxwell-Boltzmann distribution. When $E_{\{cation\},i} > \Phi_{gr}$, it is possible for the {cation} to emit an electron out of the graphene surface; a simple illustration is given in Fig. 3a to explain the proposed mechanism. For a {cation} at infinity, no interaction happens (state 1). When a {cation} with kinetic energy $E_{\{cation\},i}$ impacts the graphene surface, its kinetic energy will be converted into the internal energy of graphene due to the inelastic collision between the {cation} and the graphene (state 3). This can be verified by the presence of G$^-$ band in the Raman spectrum (Fig. 2g). Then the deformed graphene will try to release a part of the absorbed energy by releasing an electron from the bound state, and the remaining energy is used for the rebounded {cation} (state 4), which can be verified by the Raman shifts and the decreased $I_{G'}/I_G$ ratio on Raman spectrum in Fig. 2g. Since the mobility of graphene (~1000 cm$^2$V$^{-1}$s$^{-1}$) for our experiments is much higher than that of the solution, the released electrons prefer to travel across the graphene surface to the electrode instead of going into the electrolyte solution. That is how the voltage was produced by our device.

From equation (1), we assume that all the kinetic energy of the {cation} can be absorbed by graphene for releasing an electron. The induced voltage is related to $E_{\{cation\}}$, namely $V \propto E_{\{cation\}}$. $E_{min,1e} = (1/2)mv_i^2 = \Phi_{gr}$, where $\Phi_{gr}$ is 4.6 eV for our graphene samples

(Supplementary Fig. 10). For $Cu^{2+}$, $\Phi_{gr} = nE_{Cu^{2+}}$, where $n$ is the number of $Cu^{2+}$ involved in emitting an electron from graphene. The most probable velocity of $Cu^{2+}$ is ~300 m·s$^{-1}$ at room temperature, which means $n = 155$. We can reasonably define the effective cation $\{Cu^{2+}\}$ = 155 $Cu^{2+}$.

To further elaborate the proposed mechanism for the induced voltage, we carried out first-principles calculations on the interaction between graphene and $Cu^{2+}$ (see Materials and Methods in Supplementary Information). We regarded a single $Cu^{2+}$ with the same kinetic energy as the $\{Cu^{2+}\}$ in the proposed theory for simplicity. Only the interaction between the single $Cu^{2+}$ and the graphene, symmetrically above the carbon ring for simplicity, was considered. The total energy of the graphene-$Cu^{2+}$ system was calculated by the Perdew-Bure-Ernzerhof (PBE) method. The first equilibrium state located at a separation of $d_1$ between the $Cu^{2+}$ and the center of the carbon ring (Fig. 3b, corresponding to state 2 of Fig. 3a). When the distance between the $Cu^{2+}$ and graphene is larger than $d_1$, the total energy of this system keeps constant, which means no energy conversion occurs between them. When the distance approaches to $d_1$, the total energy of the system can be increased by 4.6 eV, which is equivalent to $\Phi_{gr}$. From the calculated density of states (Fig. 3c)), when 4.6 eV energy is transferred to graphene from the $Cu^{2+}$ by inelastic collision, the Fermi level of graphene shifts up by 1 eV compared to its Dirac point. This means that an electron is emitted out of the graphene.

The proposed mechanism can explain the effect of different conditions. The higher the temperature of $Cu^{2+}$ solution, the larger the kinetic energy of $\{Cu^{2+}\}$, and the higher the measured voltage (Fig. 2a and b); the higher the concentration of $Cu^{2+}$, the higher the density of {$Cu^{2+}$} on graphene plane, the more the electrons emitted out of graphene, and the higher the generated voltage (Fig. 2c). Other electrolyte solutions induced lower voltages, which can be ascribed to the difference of ionic radii, valence electrons number *n* and ion mass. Based on the proposed mechanism, we predicted that any ions or small molecules which have enough energy can excite electrons out of graphene. Based on this predication, we performed experiments use a device in DI water instead of saline solution to demonstrate it (Supplementary Fig. 11). From the results we can find that in the DI water, the induced Voc was about 9 µV, when the molecules were accelerated by Ultrasonic wave, the induced voltage will be increased by 4 times. The results can support the proposed mechanism.

We also found that asymmetric electrodes can define the current direction in the circuit. For comparison, two devices with identical electrodes, namely Au-Au and Ag-Ag, were fabricated. In such devices, it was difficult to control the current direction. That is because the excited electrons flow across graphene surface in random directions and small vibration can cause the change of the current direction (Supplementary Fig. 12a). To interpret this, a work-function tuning mechanism was proposed (Supplementary Fig. 12b).

Experiments with graphite and carbon nanotube thin film produced low voltage less than 10 µV (Supplementary Fig. 13), which can be regarded as noise. So the atomic-layer nature of graphene is crucial for the electricity generation. We also measured the output power of a typical device whose exposed area was about 10 mm × 5 mm. When a 22 Kohm resistor loaded to it, the output power reached a peak of about 1.38 µW, which means that the theoretical power density is about 73.3 KW/Kg (Supplementary Fig. 14). By putting six graphene devices in series, a $V_{oc}$ over 2.0 V was produced (Fig. 4a), which was sufficient to drive a commercial red LED (Supplementary Fig. 15). The results were clearly captured in dim background (Fig. 4b). Our experimental results present new opportunities for the development of high-performance self-charged battery technology to harvest energy from the environment.

**METHODS SUMMARY**

Monolayer graphene samples were fabricated by chemical vapor deposition (CVD) on polycrystalline copper foils using methane as the precursor[7]. They were further identified by Raman spectroscopy, scanning electron microscopy and transmission electron microscopy. A typical graphene sample of size 7 mm × 7 mm was then transferred onto $SiO_2$ (300 nm)/Si substrate. Au and Ag electrodes were deposited on either side of graphene by thermal evaporation. All the electrodes, graphene edges and substrates were sealed from exposing to the electrolyte solution. The exposed area was around 3 mm × 5 mm. *I-V* characteristic of the device exhibited good ohmic contact. Then, the device was put into the solution. The voltage generated was measured by a multimeter.

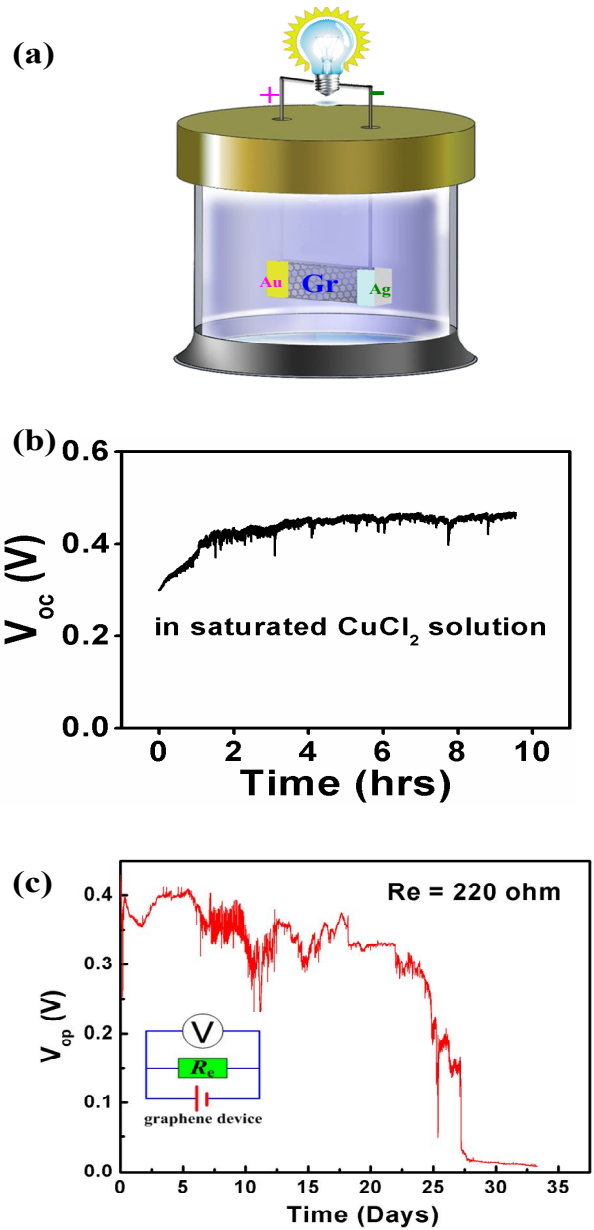

**Figure 1**. Experimental setup and output of a graphene device. a, Schematic diagram showing the experimental setup of the device with Au-Ag electrodes. b, $V_{oc}$ versus time graph in saturated $CuCl_2$ solution at room temperature. c, $V_{op}$ versus time graph of the device. Inset is an equivalent circuit of the graphene device.

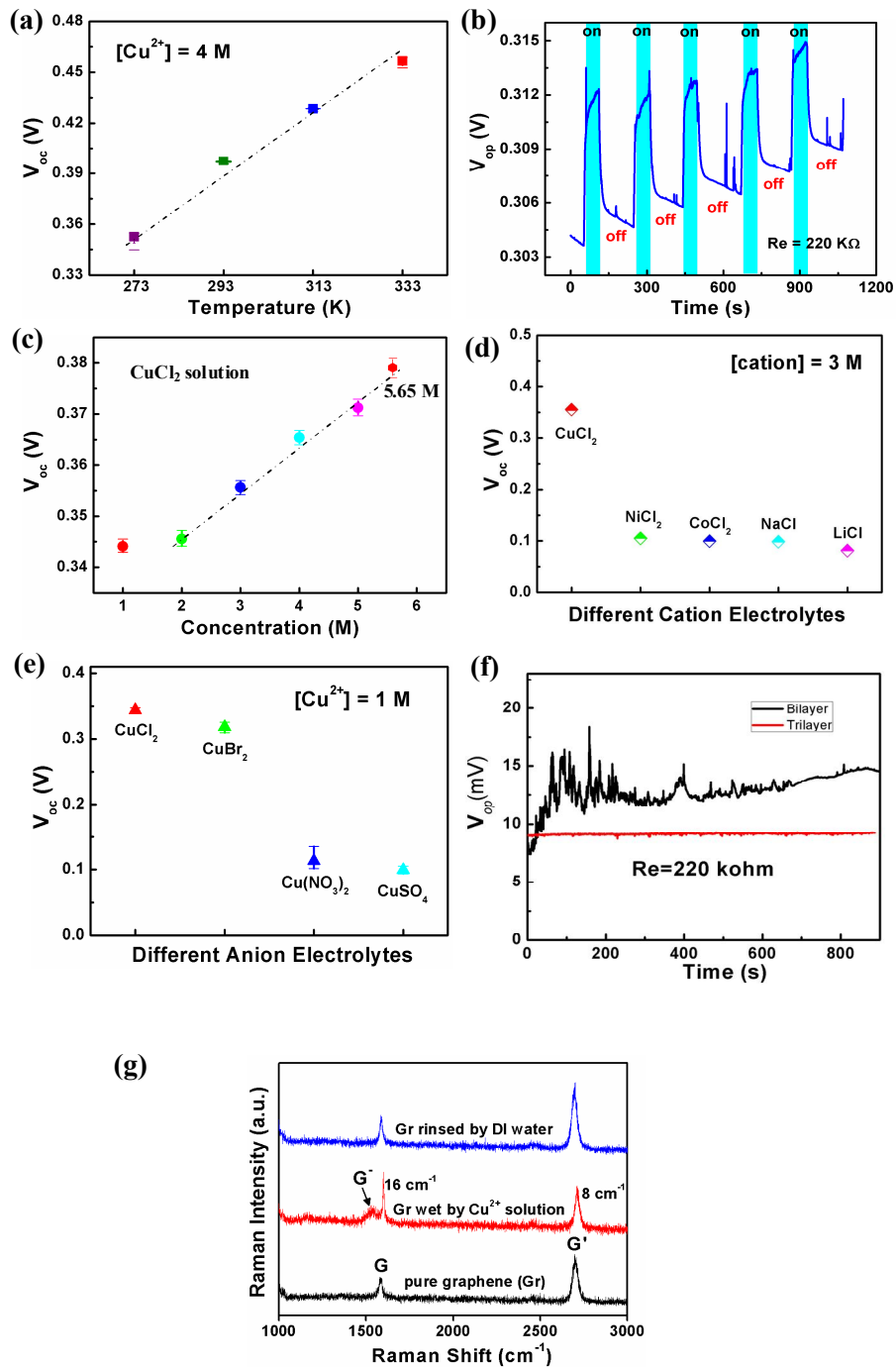

**Figure 2**. Experimental measurements of the graphene devices with Au-Ag electrodes. (a), Voc versus temperature relation in 4 M CuCl2 solution. (b), Measurement of Vop when ultrasound was intermittently turned on. (c), Voc versus concentration relation in CuCl2 solution at room temperature. (d), Measurement of Voc in 3 M electrolyte solutions with different cations at room temperature. (e), Measurement of Voc in 1 M Cu2+ solutions with different anions at room temperature. (f), $V_{op}$ of bilayer and trilayer graphene in 5.56M CuCl2 solution. (g), Raman spectra of pure graphene, graphene wet by CuCl2 solution and the sample rinsed by DI water.

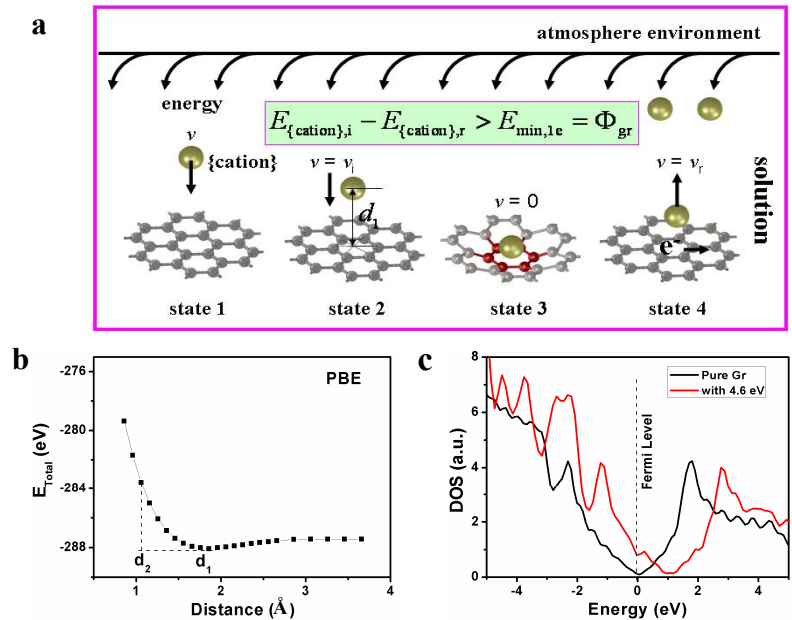

**Figure 3**. Thermal ion-graphene interaction mechanism. a, Four states for the emission of an electron: {cation} is far away from graphene surface (state 1); {cation} reaches the equilibrium location at $d_1$ with a speed of $v_i$ (state 2); impact between {cation} and graphene gives rise to the deformation of graphene (state 3); the release of an electron (state 4). b, Relationship between the total energy and the distance between the $Cu^{2+}$ and the symmetric point of the carbon ring in graphene. $d_1$ is the first equilibrium site between $Cu^{2+}$ and graphene, which is consistent with $d_1$ in state 2 of (a). At d2, 4.6eV was transferred to graphene by the interaction between $Cu^{2+}$ and graphene surface. c, Density of states of pure graphene (black) and graphene with 4.6 eV extra energy which is transformed from $Cu^{2+}$ (red). The graphene model used for simulation is 4 × 4 supercell.

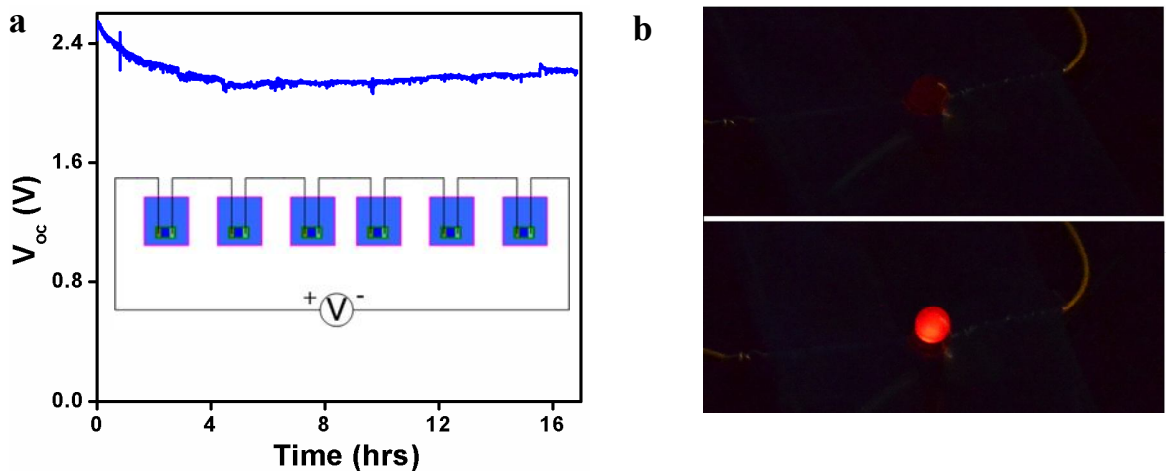

**Figure 4**. Application of the electricity generated by the devices. a, $V_{oc}$ versus time graph of six graphene samples in series in saturated $CuCl_2$ solution at room temperature. b, Images of a LED in dim background before (top) and after (bottom) it was lighted up.

# Part of the Supplementary Material:

## 2. Discussion on the Possibility of Chemical Reaction

To exclude the possibility of chemical reaction, we performed control experiments. If chemical reaction takes place, only two cases are possible: (1) chemical reaction between electrodes and solutions; (2) chemical reaction between ions and graphene. We sealed the electrodes to avoid them from exposing to the electrolyte solutions in all the experiments.

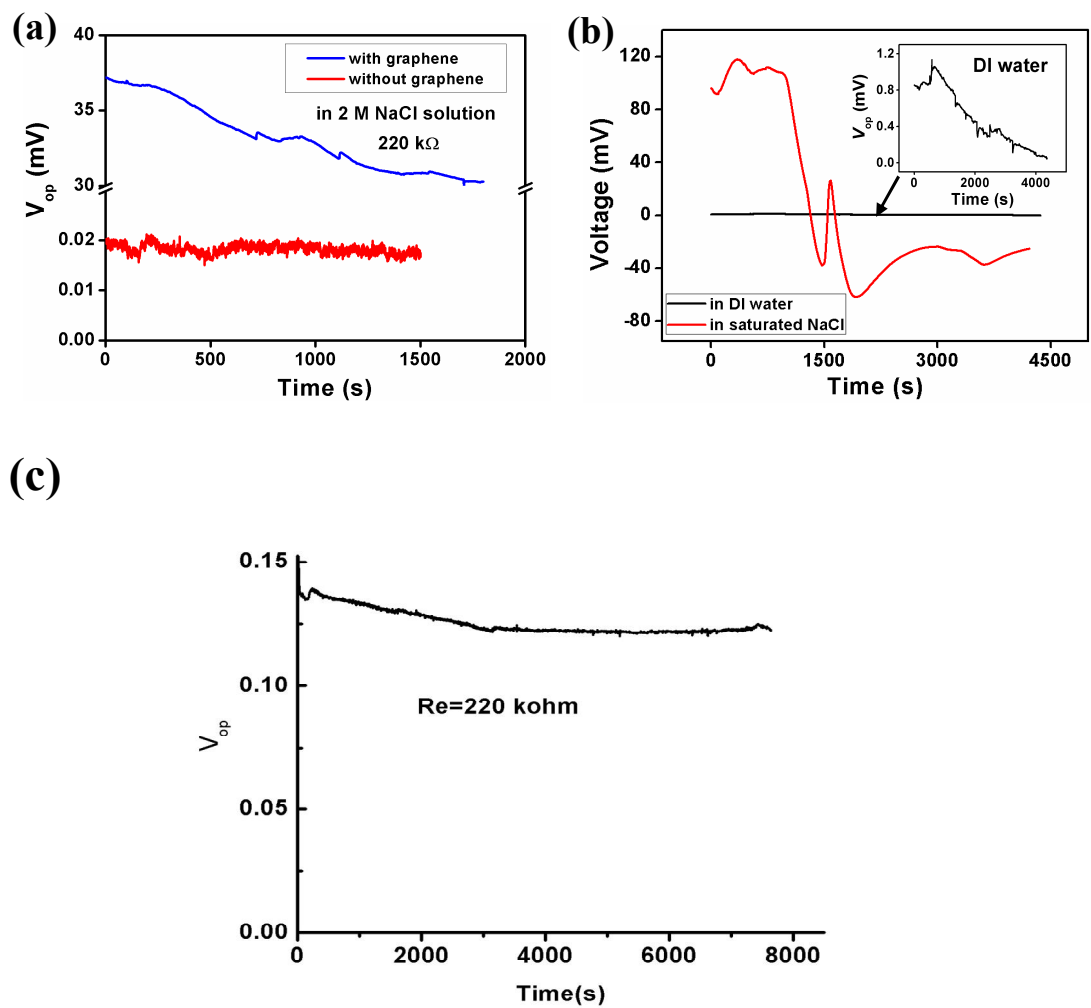

**Figure 1**. Control experiments for verifying the mechanism of the graphene electricicity generator. (a), Output of the devices (Ag-Au electrodes) with (blue) and without (red) graphene in 2 M NaCl solution. (b), Output of a device with graphite electrodes in saturated NaCl solution (red) and DI water (black). All of the devices were sealed by transparent paste, loaded with a 220 kΩ resistor in series and measured at room temperature. (c), Gold conductive wire instead of silver coated copper wire was used, graphite-graphite electrodes, monolayer graphene in 2M NaCl.

There are three arguments to exclude the possibility of chemical reaction between electrodes and solutions: (1) all the electrodes had been sealed with transparent paste before dipping into solutions, so little solution could penetrate through the paste and react with electrodes; (2) if the measured $V_{op}$ (Supplementary , Figure S1a) came from chemical reaction, there should be a pronounced voltage signal obtained from a device in the absence of graphene (only sealed Au-Ag electrodes). But the detected voltage approaches to zero (Supplementary ,black, Figure 1a); (3) we also used graphite as electrodes, which have little possibility to react with NaCl solution. However, the continuous $V_{op}$ could still be generated (Supplementary red, Figure 1b).

**Table 1**. Standard electrode potentials of cations and corresponding simple substance involved in the experiments.

| No. | Cathode (Reduction) Half-Reaction | Standard Potential E° (volts) |
| --- | --- | --- |
| (1) | $Na^+(aq) + e^- \rightleftharpoons Na(s)$ | -2.714 |
| (2) | $Cl^-(aq) \rightleftharpoons Cl_2(g) + 2e^-$ | -1.36 |
| (3) | $H_2O \rightleftharpoons O_2 + 4H^+ + 4e^-$ | -1.229 |

To exclude the possibility of chemical reaction between ions and graphene, we conducted two experiments using a sample with graphite electrodes. One in deionized (DI) water (Supplementary black, Figure 1b,), the other in saturated NaCl solution at room temperature (Supplementary red, Figure 1b,). The experimental results showed that the $V_{op}$ in deionized water is almost zero, which means that it is very difficult for the chemical reaction between graphene and water to happen. We can find that if there are any chemical reactions between graphene and the ions ( $Na^+$, $Cl^-$, $H_3O^+$), it is easier for $H_3O^+$ to be reacted than other ions (Supplementary Table S1,). So, the $V_{op}$ in NaCl solution was not resulted from chemical reaction (Supplementary Figure 1b).

We performed the third group experiments with gold wire instead of copper wire to avoid the chemical recation between solution and the conductive wires. The electrodes wad made by graphite and the solution is 2M NaCl. The results in Figure S1c showed that the effect of harvest electricity is from the chemical reaction between conductive wire and solution.

In conclusion, we could not find any evidences that support the opinion that the induced voltage came from chemical reaction. The mechanism for electricity generation by graphene in solution is a pure physical process, which is discussed in details in the text.

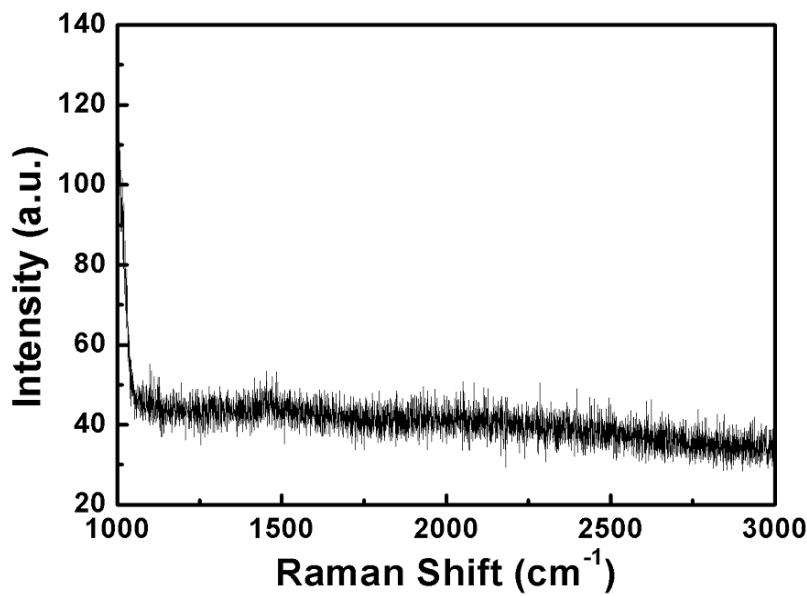

**Fig 4, The raman spectrum of the sample after being measured for 33 days.**

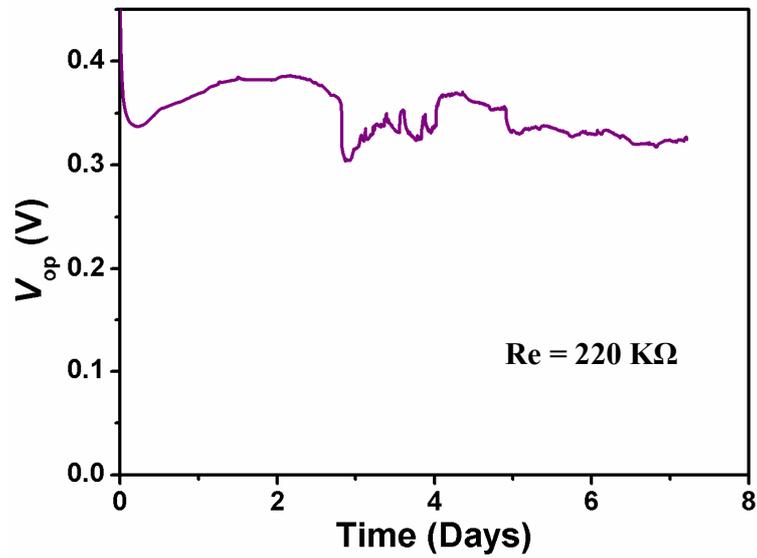

**Figure 5.** $V_{op}$ versus time relation of the device with a 220 KΩ resistor loaded to the circuit in saturated CuCl$_2$ solution.

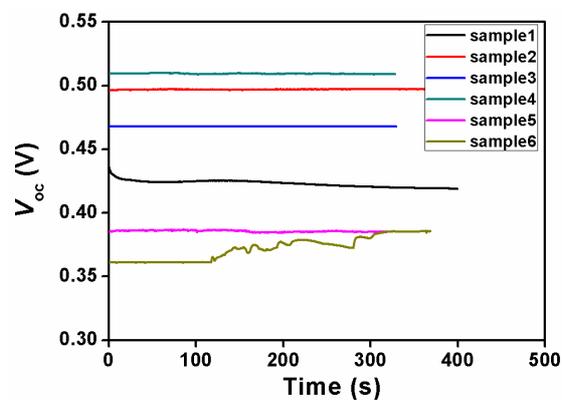

**Figure 6**. $V_{oc}$ of six graphene devices in saturated CuCl$_2$ solution at room temperature. The measured voltages vary from 0.36 to 0.51 V.

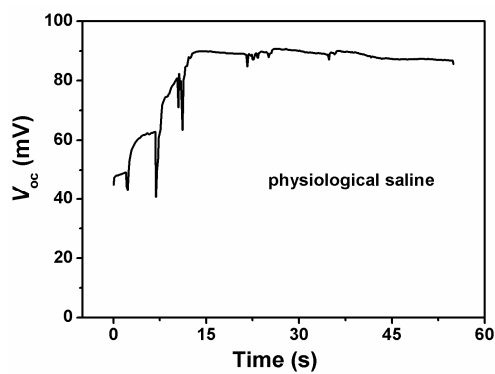

**Figure 7.** $V_{oc}$ versus time line of a device in physiological saline.

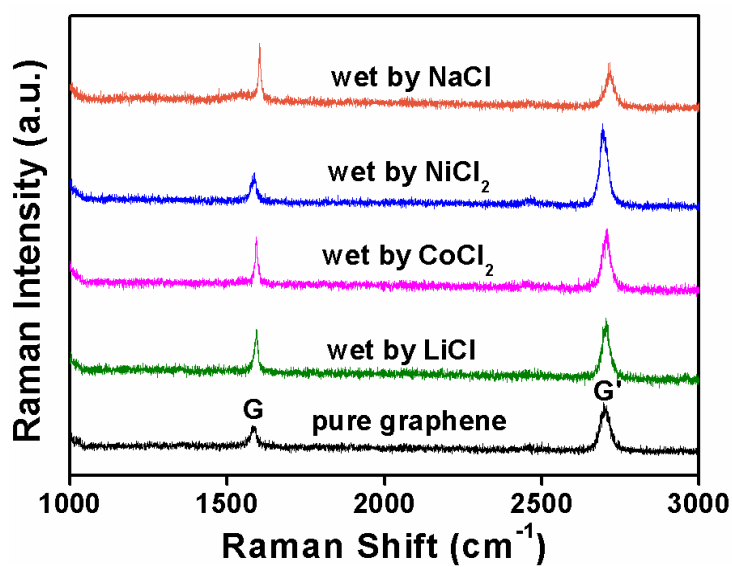

**Figure 9**. Raman spectra of pristine graphene and graphene samples dipped by different solutions: $CoCl_2$, $NiCl_2$, NaCl and LiCl.

As can be seen from Raman spectrum of pure graphene (bottom, Figure S6, Supporting Information), the two most obvious/apparent features are a G band at ~1580 cm$^{-1}$ and a G' band at ~2700 cm$^{-1}$. A weak G⁻ band appears in Raman spectra of graphene wet by electrolyte solutions and the decrease of the $I_{G'}/I_G$ ratios were observed after dipping in all solutions except NiCl$_2$.

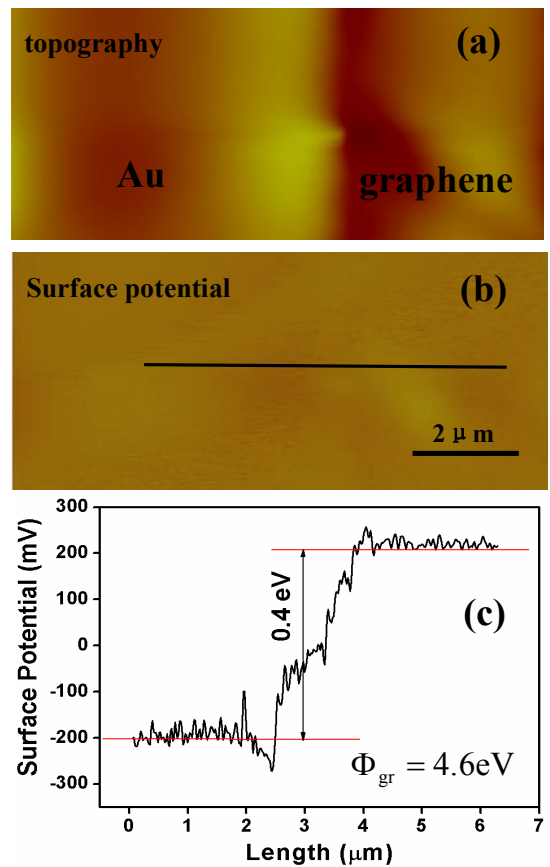

**Figure 10**. Work function of graphene measured by conductive atomic force microscopy. (a), Topography of graphene. (b), Corresponding surface potential of graphene. (c), Potential profile recorded in the sample, which is marked as a black line in (b).

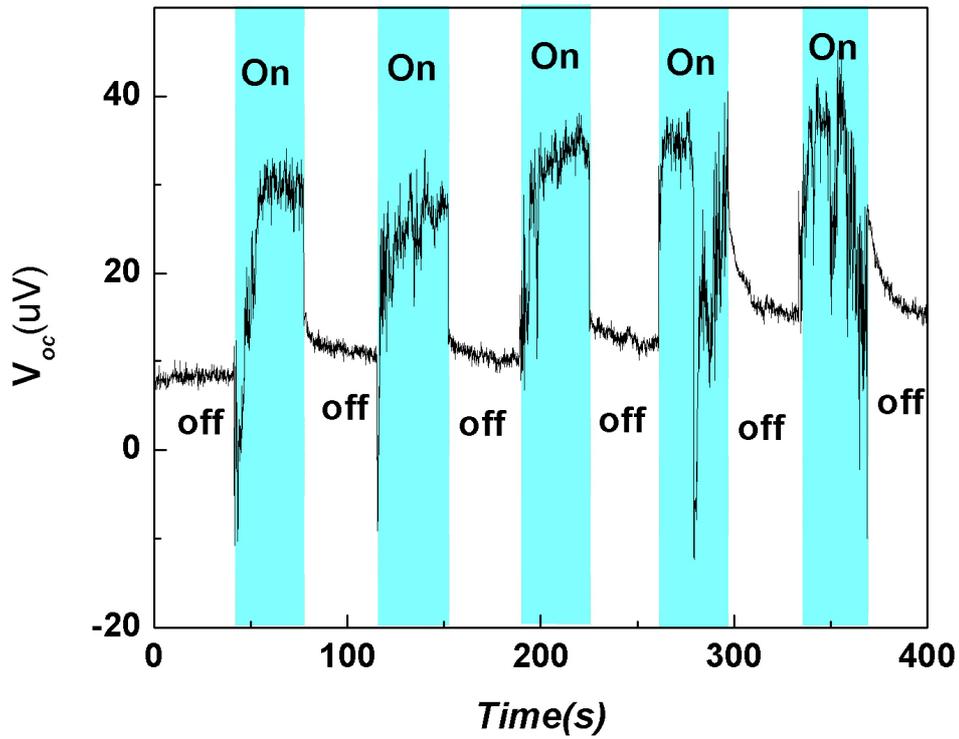

Fig 11, Voc of a graphene sample in DI water with Ultrasonic wave treated it.

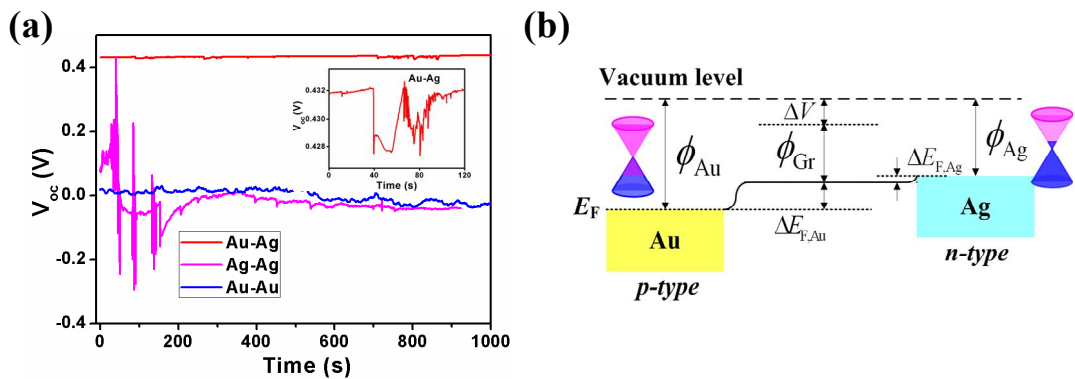

**Figure 12.** Work function tuning mechanism of the graphene device with asymmetric electrodes. a, $V_{op}$ of the graphene devices with Au-Ag, Au-Au and Ag-Ag as the electrodes respectively under small vibration. b, Energy level diagram for graphene and metal contact

exhibiting a work-function tuning mechanism. The black dashed line denotes the Fermi level, the black solid line represents the energy at the Dirac point of the graphene, the yellow rectangle is the Au contact, and the skyblue rectangle is the Ag contact. The red and blue crosses show the linear dispersion around Dirac point. $\phi_{Au}$, $\phi_{Ag}$ and $\phi_{Gr}$ are the work function of Au, Ag and graphene, respectively. $\Delta\phi$ is the total built-in potential difference. $\Delta E_{F,Au}$ and $\Delta E_{F,Ag}$ are the differences between the Dirac-point and Fermi-level energies in the Au- and Ag- graphene.

The current direction can be determined for the device with Au-Ag as asymmetric electrodes (Supplementary ,red, Figure 11a).This is because work function of gold (5.0 eV) is larger than that of graphene (4.6 eV), while work function (4.26 eV) of silver is smaller than that of graphene. Therefore, we present a work-function tuning mechanism based on the overall experimental results (Figure 11b).

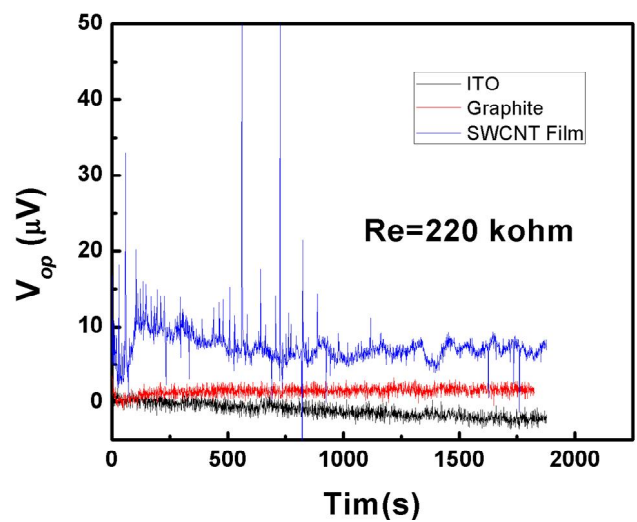

**Figure 13.** $V_{oc}$ of the devices based on arrays of CNTs (~200 nm thickness) and graphite with Au-Ag electrodes.

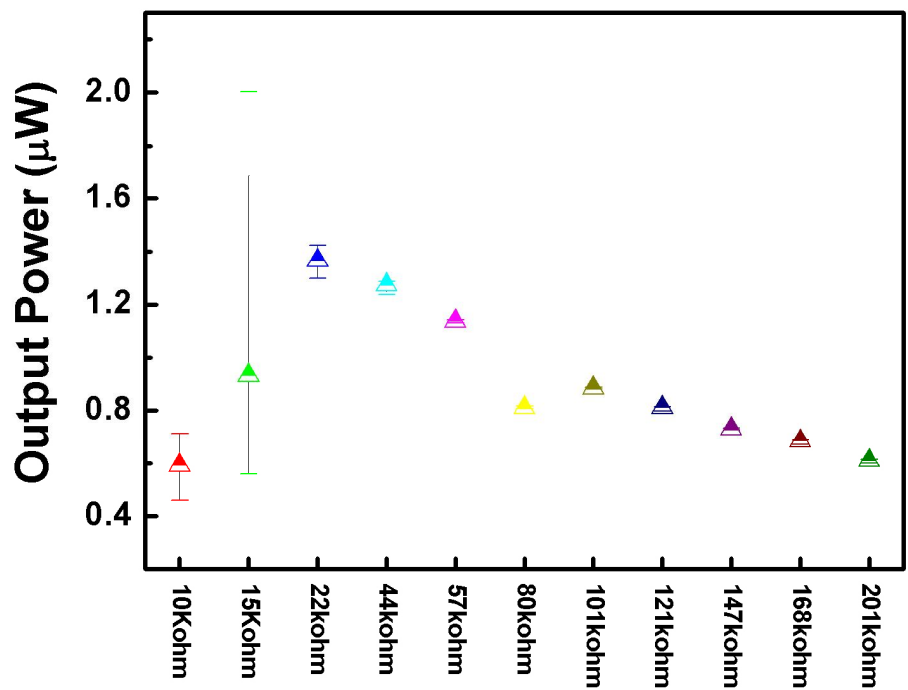

**Figure 14. Output power of a typical device in 5M CuCl2.**

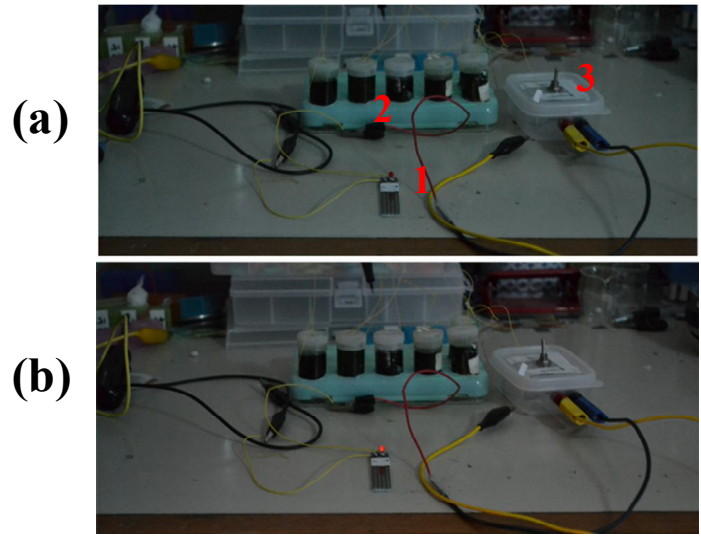

**Figure 15**. Experimental setup of six graphene devices connected with a commercial LED (SL-3ER2SD00, 625-635 nm) before (a) and after (b) it was lighted up.

In the above figure, 1 is our LED, 2 is six devices in series and 3 is a switch.